\documentclass[sigconf]{acmart}

\renewcommand\footnotetextcopyrightpermission[1]{} %
\usepackage{bbm}
\usepackage{balance}
\usepackage{cleveref}
\usepackage{multirow}
\usepackage{graphicx}
\usepackage{enumitem}
\usepackage{xcolor}

\usepackage{algorithm,algpseudocode}%

\usepackage{bm}
\usepackage{color}
\usepackage{apxproof}
\usepackage{gensymb}
\usepackage{pifont}%

\usepackage{xspace}
\usepackage{mathtools}

\newcommand{\squishlist}{
 \begin{list}{$\bullet$}
  { \setlength{\itemsep}{0pt}
     \setlength{\parsep}{3pt}
     \setlength{\topsep}{3pt}
     \setlength{\partopsep}{0pt}
     \setlength{\leftmargin}{1.5em}
     \setlength{\labelwidth}{1em}
     \setlength{\labelsep}{0.5em} } }

\newcommand{\squishlisttwo}{
 \begin{list}{$\bullet$}
  { \setlength{\itemsep}{0pt}
     \setlength{\parsep}{0pt}
    \setlength{\topsep}{0pt}
    \setlength{\partopsep}{0pt}
\setlength{\leftmargin}{2em}
\setlength{\labelwidth}{1.5em}
\setlength{\labelsep}{0.5em} } }

\newcommand{\squishend}{
\end{list}  }

\AtBeginDocument{%
  \providecommand\BibTeX{{%
    \normalfont B\kern-0.5em{\scshape i\kern-0.25em b}\kern-0.8em\TeX}}}

\newcommand{\eg}{\emph{e.g.}}
\newcommand{\ie}{\emph{i.e.}}

\begin{document}

\newcommand{\cl}[1]{{\textcolor{red}{[Liu: #1]}}}

\title{Mamba4Rec: Towards Efficient Sequential Recommendation \\ with Selective State Space Models}

\author{Chengkai Liu}
\affiliation{
  \institution{Texas A\&M University}
  \city{College Station, Texas}
  \country{USA}}
\email{liuchengkai@tamu.edu}

\author{Jianghao Lin}
\affiliation{
  \institution{Shanghai Jiao Tong University}
  \city{Shanghai}
  \country{China}}
\email{chiangel@sjtu.edu.cn}

\author{Jianling Wang}
\affiliation{
  \institution{Texas A\&M University}
  \city{College Station, Texas}
  \country{USA}}
\email{jwang713@tamu.edu}

\author{Hanzhou Liu}
\affiliation{
  \institution{Texas A\&M University}
  \city{College Station, Texas}
  \country{USA}}
\email{hanzhou1996@tamu.edu}

\author{James Caverlee}
\affiliation{
  \institution{Texas A\&M University}
  \city{College Station, Texas}
  \country{USA}}
\email{caverlee@tamu.edu}

\renewcommand{\shortauthors}{Liu, et al.}

\begin{abstract}

Sequential recommendation aims to estimate the dynamic user preferences and sequential dependencies among historical user behaviors. 
Although Transformer-based models have proven to be effective for sequential recommendation, they suffer from the inference inefficiency problem stemming from the quadratic computational complexity of attention operators, especially for long behavior sequences. 
Inspired by the recent success of state space models (SSMs), we propose Mamba4Rec, which is the first work to explore the potential of selective SSMs for efficient sequential recommendation. 
Built upon the basic Mamba block which is a selective SSM with an efficient hardware-aware parallel algorithm, we design a series of sequential modeling techniques to further promote model performance while maintaining inference efficiency. Through experiments on public datasets, we demonstrate how Mamba4Rec effectively tackles the effectiveness-efficiency dilemma, outperforming both RNN- and attention-based baselines in terms of both effectiveness and efficiency. 
The code is available at \url{https://github.com/chengkai-liu/Mamba4Rec}.

\end{abstract}


\maketitle

\section{Introduction}
Personalized online services rely heavily on sequential recommender systems to capture dynamic user preferences~\cite{clickprompt,wang2021survey,lu2015recommender,lin2023map}. 
These systems aim to predict users' future interactions by effectively modeling sequential dependencies among historical user behaviors. 
Earlier neural approaches adopt convolutional neural networks (CNNs)~\cite{tang2018personalized} and recurrent neural networks (RNNs)~\cite{hidasi2015session, li2017neural}, which have pioneered the application of neural networks in sequential recommendation yet suffer from the catastrophic forgetting issue~\cite{schak2019study,kirkpatrick2017overcoming}.
More recently, researchers have adopted the  Transformer~\cite{vaswani2017attention} as the backbone module and leverage the self-attention mechanism to model the dynamic user behavior sequences~\cite{kang2018self, yuan2022multi}.
Despite their remarkable performance, these attention-based methods generally suffer from the inference inefficiency problem due to the quadratic computational complexity inherent in attention operators, especially for long user behavior sequences or large-scale datasets~\cite{tay2020long, tay2022efficient}. This tradeoff between performance and efficiency has hindered their practical application.

Recently, state space models (SSMs)~\cite{gu2021efficiently}, such as S4~\cite{gu2021efficiently}, S5~\cite{smith2022simplified} and Mega~\cite{ma2022mega}, have been widely adopted as alternatives to RNNs, CNNs, and Transformers in various natural language tasks. 
These models inherently possess inference efficiency due to their recurrent nature, and meanwhile enjoy strong performance to handle long-range dependencies through structured state matrices. 
Moreover, a recently proposed SSM variant (\ie, Mamba~\cite{gu2023mamba}) further introduces the input-dependent selective mechanism with efficient hardware-aware design to tackle the data-independent issue of prior SSMs.
This allows the model to selectively extract essential knowledge and filter out irrelevant noise according to the input data, thus leading to superior sequential modeling performance.
These advancements provide the potential to address the tradeoff between recommendation performance and inference efficiency, positioning SSMs as a core operator for sequential recommendation.

In this work, we provide an in-depth investigation of adopting selective SSMs for sequential recommendation, introducing \textbf{Mamba4Rec}, the first model to leverage the power of selective SSMs for efficient sequential recommendation. 
Built upon the basic Mamba block~\cite{gu2023mamba}, we incorporate and discuss a series of techniques (\eg, position embedding~\cite{vaswani2017attention}, residual connection~\cite{he2016deep}, layer normalization~\cite{ba2016layer}, and position-wise feed-forward network~\cite{vaswani2017attention}), which supplements the system non-linearity, stabilizes the training dynamics, and thereby further promotes the sequential modeling capability.
The main contributions of this paper are as follows:
\begin{itemize}[leftmargin=10pt]
    \item We are the first to explore the potential of selective SSMs for sequential recommendation, to address the dilemma of recommendation performance and inference efficiency.
    \item By understanding the impact of the basic Mamba block, including its incorporation of normalization and feed-forward networks for sequential recommendation, we propose Mamba4Rec, which further improves sequential modeling capability without sacrificing inference efficiency.
    \item Experiments on public datasets demonstrate the superiority of our proposed Mamba4Rec compared with RNN- and attention-based baselines in terms of both effectiveness and efficiency.
\end{itemize}

\section{Preliminaries}

In sequential recommendation, let $\mathcal{U}=\left\{u_1, u_2, \ldots, u_{|\mathcal{U}|}\right\}$ denote the user set, $\mathcal{V}=\left\{v_1, v_2, \ldots, v_{|\mathcal{V}|}\right\}$ denote the item set, and $\mathcal{S}_u=\left\{v_1, v_2, \ldots, v_{n_u}\right\}$ denote the chronologically ordered interaction sequence for user $u \in \mathcal{U}$, where $n_u$ is the the length of the sequence. Given the interaction history $\mathcal{S}_u$, the task is to predict the next interacted item, denoted as $v_{n_{u}+1}$ for user $u$.

\smallskip
\noindent\textbf{State Space Models (SSM).} The State Space Model (SSM) is a sequence modeling framework based on linear ordinary differential equations.
It maps an input sequence $x(t) \in \mathbb{R}^{D}$ to the output sequence $y(t) \in \mathbb{R}^N$ through the latent state $h(t) \in \mathbb{R}^N$:
\begin{equation}
    \label{eq:ssm}
    \begin{aligned}
    h^{\prime}(t) & =\boldsymbol{A} h(t)+\boldsymbol{B} x(t), \\
    y(t) & =\boldsymbol{C} h(t),
    \end{aligned}
\end{equation}
where $\boldsymbol{A} \in \mathbb{R}^{N \times N}$ and $\boldsymbol{B}, \boldsymbol{C} \in \mathbb{R}^{N \times D}$ are learnable matrices. 
To model discrete sequences instead of continuous functions, it must be discretized by a step size $\Delta$. Alternatively, the discretized SSM can be expressed as follows:
\begin{equation}
\begin{aligned}
h_t & =\overline{\boldsymbol{A}} h_{t-1}+\overline{\boldsymbol{B}} x_t, \\
y_t & =\boldsymbol{C} h_t,
\end{aligned}
\end{equation}
where $\overline{\boldsymbol{A}}=\exp (\Delta \boldsymbol{A})$ and $\overline{\boldsymbol{B}}=(\Delta \boldsymbol{A})^{-1}(\exp (\Delta \boldsymbol{A})-\boldsymbol{I}) \cdot \Delta \boldsymbol{B}$. 
After converting the parameters from a continuous form $(\Delta, \boldsymbol{A}, \boldsymbol{B}, \boldsymbol{C})$ to a discrete form $(\overline{\boldsymbol{A}}, \overline{\boldsymbol{B}}, \boldsymbol{C})$, the model can be computed in a linear recurrence way~\cite{gu2021combining}, which enhances computational efficiency. 
Deriving from the vanilla SSM, the structured state space model (S4)~\cite{gu2021efficiently} imposes structure on the state matrix $\boldsymbol{A}$ with HiPPO~\cite{gu2020hippo} initialization to further improve the long sequence modeling.

Building upon S4, Mamba~\cite{gu2023mamba} introduces an input-dependent selection mechanism and leverages a hardware-aware parallel algorithm in recurrent mode. This combined approach empowers Mamba to capture contextual information effectively, particularly in long sequences, while maintaining computational efficiency. As a linear-time sequence model, Mamba achieves Transformer-quality performance with better efficiency, especially on long sequences.

\section{Mamba4Rec}

In this section, we introduce our proposed framework, Mamba4Rec. We begin with a high-level overview of the framework, followed by an exploration of its technical components. We describe how Mamba4Rec constructs a sequential recommendation model through an embedding layer, selective state space models, and a prediction layer. Additionally, we discuss the key components widely used in sequential recommendation, such as positional embeddings~\cite{vaswani2017attention}, feed-forward networks, dropout~\cite{srivastava2014dropout}, and layer normalization~\cite{ba2016layer}.

\subsection{Framework Overview}

\begin{figure}[ht]
    \centering
    \includegraphics[scale=0.46]{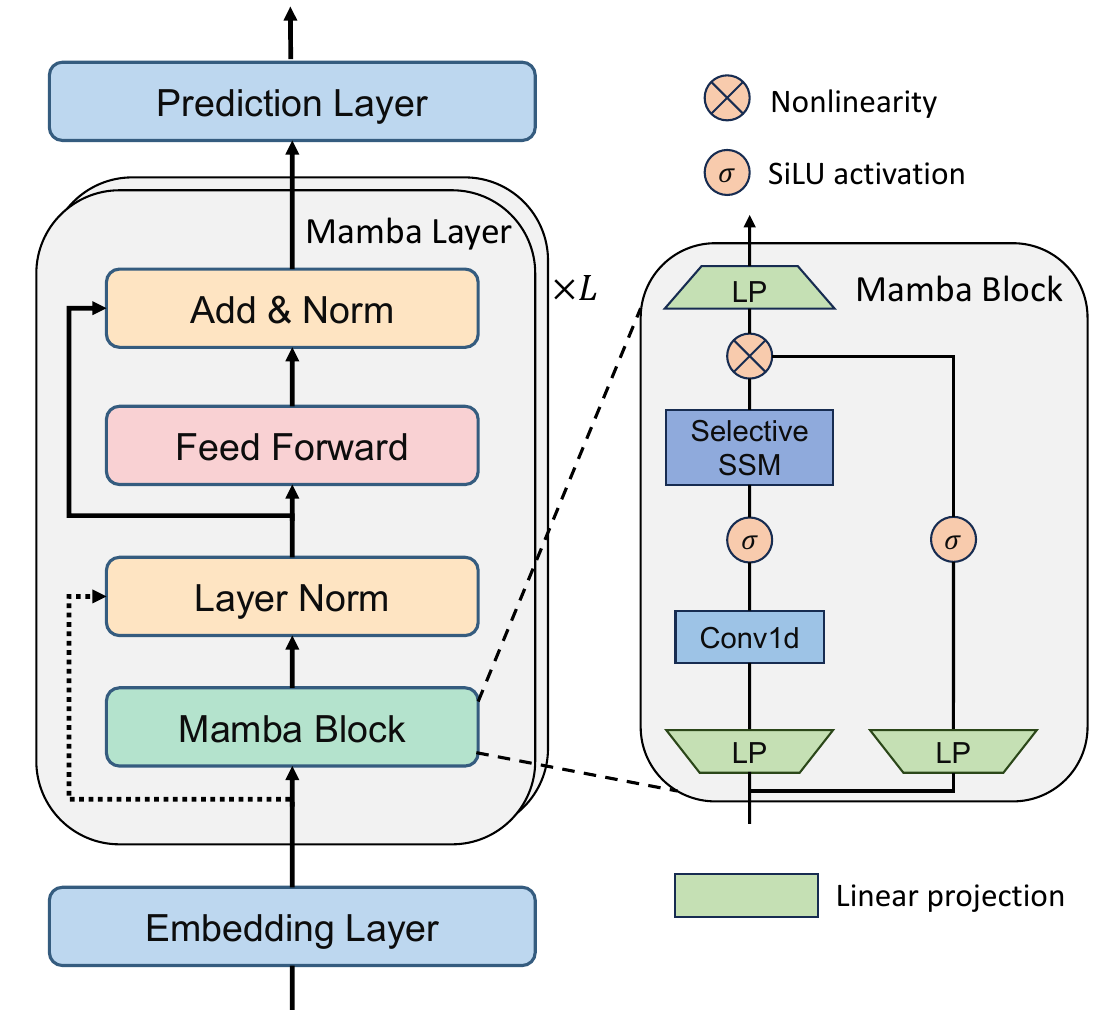} 
    \caption{The overview of Mamba4Rec.}
    \label{fig:framework}
\end{figure}

As illustrated in Figure~\ref{fig:framework}, the proposed Mamba4Rec is a sequential recommendation model that leverages selective state space models through Mamba blocks. The core element of Mamba4Rec is the Mamba layer, which combines a Mamba block with a position-wise feed-forward network. Mamba4Rec offers flexibility in its architecture: while it can be stacked with multiple Mamba layers, a single Mamba layer often proves sufficient. Each layer comprises a Mamba block followed by a feed-forward network, enabling the model to capture both item-specific information and sequential context effectively from the user's interaction history.

\subsection{Embedding Layer}

Similar to existing models, our approach utilizes an embedding layer to map item IDs to a high-dimensional space. This embedding layer uses a learnable embedding matrix $\boldsymbol{E} \in \mathbb{R}^{|\mathcal{V}| \times D}$, where $D$ is the embedding dimension. By applying the embedding layer to the input item sequence $\mathcal{S}_u$, we obtain the initial item embeddings $\mathcal {S}_u \boldsymbol E$. To enhance robustness and prevent overfitting, we incorporate both embedding dropout and layer normalization after retrieving the embeddings:
\begin{equation}
\begin{aligned}
 H & =\operatorname {LayerNorm}\left(\operatorname {Dropout}\left( \mathcal {S}_u \boldsymbol E\right)\right)\in \mathbb{R}^{n_u \times D}.
\end{aligned}
\label{eq:layernorm}
\end{equation}
During the training and inference stage, we merge multiple samples into one mini-batch, resulting in the input data $x\in\mathbb{R}^{B\times L\times D}$ for the subsequent Mamba block, where $B$ is the batch size and $L$ is padded sequence length.

\vspace{1.5ex}\noindent \textbf{Positional Embedding.} We investigate the impact of positional embeddings on our model's performance. This includes evaluating learnable positional embeddings as used in~\cite{kang2018self, sun2019bert4rec}.  These embeddings are typically injected into the input item embedding before layer normalization. However, unlike Transformer-based models that rely on positional embeddings to capture the sequence order, state space models inherently handle sequential information through their recurrent nature. While theoretically unnecessary, we analyze the effect of positional embeddings on our model's behavior in our experiments.

\subsection{Mamba Layer}

\begin{table*}[t]
	\small
 \vspace{-5pt}
	\caption{Recommendation performance. The best results are bold, and the second-best are underlined. 
  The symbol $*$ indicates a statistically significant improvement of Mamba4Rec over the best baseline with $p$-value < 0.01.}
  \vspace{-10pt}
	\label{tab:performance}
	\renewcommand\arraystretch{1.1}
	\begin{center} {	
	\begin{tabular}{cc|ccc|ccc|ccc}
		\toprule
		\multicolumn{2}{c}{\multirow{2}{*}{\textbf{Method}}} & \multicolumn{3}{c}{\textbf{MovieLens-1M}} & \multicolumn{3}{c}{\textbf{Amazon-Beauty}} & \multicolumn{3}{c}{\textbf{Amazon-Video-Games}} \cr
  
		\cmidrule(lr){3-5}\cmidrule(lr){6-8}\cmidrule(lr){9-11} & 
  &\textbf{HR@10} & \textbf{NDCG@10} & \textbf{MRR@10} &\textbf{HR@10} & \textbf{NDCG@10} & \textbf{MRR@10}&\textbf{HR@10} & \textbf{NDCG@10} & \textbf{MRR@10} \\
  
  \hline	
    &BPR-MF & 0.1702 & 0.0891 & 0.0645 & 0.0576 & 0.0322 & 0.0245 & 0.0686 & 0.0386 & 0.0295 \\
    &Caser    & 0.2399 & 0.1283 & 0.0944 & 0.0582 & 0.0318 & 0.0239 & 0.0755 & 0.0425 & 0.0324 \\
    &NARM    & 0.2735 & 0.1506 & 0.1132 & 0.0627 & 0.0347 & 0.0262 & 0.1032 & 0.0530 & 0.0379 \\
    &GRU4Rec & 0.2934 & 0.1642 & 0.1249 & 0.0606 & 0.0332 & 0.0249 & 0.1030 & 0.0536 & 0.0380 \\
    &SASRec & 0.2977 & 0.1687 & 0.1294 & \textbf{0.0847} & \underline{0.0425} & \underline{0.0296} & \textbf{0.1168} & \underline{0.0571} & \underline{0.0390}\\
    &BERT4Rec & 0.2958 & 0.1674 & 0.1275 & 0.0760 & 0.0393 & 0.0282 & 0.1053 & 0.0538 & 0.0381\\
    &LRURec & \underline{0.3057} & \underline{0.1772} & \underline{0.1380} & 0.0759 & 0.0393 & 0.0280 & 0.1134 & 0.0560 & 0.0386\\
    \hline
    &Mamba4Rec & \textbf{0.3121}$^*$ & \textbf{0.1822}$^*$ &  \textbf{0.1425}$^*$ & \underline{0.0812} & \textbf{0.0451}$^*$ &\textbf{0.0362}$^*$ & \underline{0.1152} & \textbf{0.0603}$^*$ & \textbf{0.0438}$^*$\\
		\bottomrule
		\end{tabular}}
  \vspace{-5pt}
	\end{center}

\end{table*}

While the Mamba block~\cite{gu2023mamba} has potential as a sequence-to-sequence model, we leverage its strengths for sequential recommendation tasks. We introduce the Mamba layer, which incorporates both the Mamba block and a position-wise feed-forward network. Instead of simply stacking Mamba blocks, we use the Mamba layer as the fundamental building block for our architecture.

\vspace{1.5ex}\noindent \textbf{Mamba Block.} 
As detailed in Algorithm~\ref{alg:mamba}, the Mamba block operates on an input $H_i \in \mathbb{R}^{B \times L \times D}$ with dimensions of batch size $B$, sequence length $L$, and hidden dimension $D$. It begins by performing linear projections on the input $H_i$ with an expanded hidden dimension of $D$ to obtain $H_x$ and $H_z$. These projections are then processed through a 1D convolution and a SiLU~\cite{elfwing2018sigmoid} activation. The core of the block involves a selective state space model with parameters discretized based on the input. This discretized SSM, along with $H'_x$, generates the state representation $H_y$. Finally, $H_y$ is combined with a residual connection from $H_z$ after applying SiLU, and a final linear projection delivers the final output $H_o$. Overall, the Mamba block leverages input-dependent adaptations and selective SSM to process sequential information effectively. The parameters of the Mamba block consist of an SSM state expansion factor $N$, a kernel size $K$ for the convolution, and a block expansion factor $E$ for input and output linear projections. Selecting the expansion factors $N$ and $E$ involves a tradeoff between (capturing complex relationships) and training efficiency (higher $N, E$ increase computational cost).

The Mamba block addresses the prevalent effectiveness-efficiency dilemma existing in RNNs and Transformers. The recurrent SSM in the Mamba block operates in linear time with respect to the sequence length, making it as efficient as RNNs for inference, achieved by discretized parameters. Additionally, the selection mechanism makes the parameters of SSM input-dependent. This enables Mamba to selectively remember or forget information, crucial for capturing complex dependencies within sequential data. With the structured state matrix, it outperforms self-attention blocks in modeling long user behavioral sequences.

\begin{algorithm}
\caption{Mamba Block with Selective SSM}
\label{alg:mamba}
\begin{algorithmic}[1]
    \Statex \textbf{Input:} $H_i:(B, L, D)$
    \Statex \textbf{Output:} $H_o:(B, L, D)$
    \State $H_x, H_z:(B, L, ED) \leftarrow \operatorname{Linear}(H_i)$
    \Comment{Linear projections of input}
    \State $H'_x:(B, L, ED) \leftarrow \operatorname{SiLU}(\operatorname{Conv1d}(H_x))$
    
    \State $\boldsymbol{A}:(D, N) \leftarrow \text{Parameter}$
    \Comment{Structured state matrix}
    \State $\boldsymbol{B, C}:(B, L, N) \leftarrow \operatorname{Linear}(H'_x),  \operatorname{Linear}(H'_x)$
    \State $\Delta:(B, L, D) \leftarrow \operatorname{Softplus}(\text{Parameter} + \operatorname{Broadcast}(\operatorname{Linear}(H'_x)))$
    \State $\overline{\boldsymbol{A}}, \overline{\boldsymbol{B}}:(B, L, D, N) \leftarrow \text{discretize}(\Delta, \boldsymbol{A}, \boldsymbol{B})$
    
    \Comment{Input-dependent parameters and discretization}
    \State $H_y:(B, L, ED) \leftarrow \text{SelectiveSSM}(\overline{\boldsymbol{A}}, \overline{\boldsymbol{B}}, \boldsymbol{C})(H'_x)$
    \State $H_{y'}:(B, L, ED) \leftarrow H_y \otimes \operatorname{SiLU}(H_z)$
    \State $H_o:(B, L, D) \leftarrow \operatorname{Linear}(H_{y'})$
    \Comment{Final linear projection}
    \State \textbf{return} $H_o$ 
\end{algorithmic}
\end{algorithm}

\vspace{1.5ex}\noindent \textbf{Position-wise Feed-Forward Network.} We leverage a position-wise feed-forward network (PFFN) to improve the modeling of user actions in the hidden dimension:
\begin{equation}
\operatorname{PFFN}(H)= \operatorname{GELU}\left(H\boldsymbol{W}^{(1)}+\boldsymbol{b}^{(1)}\right)\boldsymbol{W}^{(2)}+\boldsymbol{b}^{(2)},
\end{equation}
where $\boldsymbol{W}^{(1)} \in \mathbb{R}^{D \times 4 D}, \boldsymbol{W}^{(2)} \in \mathbb{R}^{4 D \times D}, \boldsymbol{b}^{(1)} \in \mathbb{R}^{4 D}$ and $\boldsymbol{b}^{(2)} \in \mathbb{R}^{D}$ are parameters of two dense layers and we use GELU~\cite{hendrycks2016gaussian} activation. The PFFN aims to capture complex patterns and interactions within the sequential data by applying two non-linear transformations using dense layers with the activation function. 
Additionally, to enhance model robustness and prevent overfitting, we utilize dropout and layer normalization, similar to Equation~\ref{eq:layernorm}, after each Mamba block and feed-forward network within the Mamba layers. This helps to regularize the model and accelerate training convergence.

\vspace{1.5ex}\noindent \textbf{Stacking Mamba Layers}
We also explore using stacked Mamba layers in sequential recommendation. While deeper networks with multiple layers are not inherently beneficial for performance, residual connections~\cite{he2016deep} play a crucial role in facilitating the propagation of low-level features to higher layers throughout the multi-layer architecture. Recognizing the potential benefits of stacked Mamba layers, especially on long sequences, we explore two configurations for Mamba layers that balance effectiveness and efficiency:
\begin{itemize}[leftmargin=10pt]
    \item \textbf{Single Layer}: When using only one Mamba layer, we apply layer normalization without residual connections. A single Mamba layer demonstrates comparable effectiveness to stacking Mamba layers while achieving higher efficiency in our tasks.
    \item \textbf{Stacked Layers}: For stacked Mamba layers, residual connections become essential and enable the propagation of the last visited item's embedding to the final layer.
\end{itemize}
To explore the tradeoff between effectiveness and efficiency, we further investigate the potential of stacked layers in experiments.

\subsection{Prediction Layer}
In the prediction layer, we adopt the approach of SASRec, using the last item embedding to generate the final output prediction scores:
\begin{equation}
    \hat y=\operatorname{Softmax}\left(h \boldsymbol{E}^\top \right) \in \mathbb{R}^{|\mathcal{V}|},
\end{equation}
where $h \in \mathbb{R}^D$ is the last item embedding from the Mamba layer. $\hat y \in \mathbb{R}^{|\mathcal{V}|}$ represents the probability distribution over the next item in the item set $\mathcal{V}$.

\section{Experiments}

\subsection{Experimental Setup}

\noindent \textbf{Datasets.} 
We conduct experiments and evaluate our models using three real-world datasets:
\begin{itemize}[leftmargin=10pt]
    \item \textbf{MovieLens-1M}~\cite{harper2015movielens}: A benchmark dataset contains about 1 million movie ratings from users.
    \item \textbf{Amazon-Beauty} and \textbf{Amazon-Video-Games}~\cite{mcauley2015image}: Two datasets contain product reviews and ratings for the ``Beauty'' and ``Video Games'' categories on Amazon.
\end{itemize}
For each user, we construct an interaction sequence by sorting their interaction records based on timestamps. We filter users and items with less than 5 recorded interactions, following the setting in previous works~\cite{kang2018self}. The detailed statistics for each dataset after preprocessing are summarized in Table~\ref{tab:dataset}.

\begin{table}[ht]
\normalsize
\tabcolsep=5.5pt
  \centering
  \caption{Statistics of the experimented datasets.}
  \vspace{-7pt}
  \label{tab:dataset}
  \resizebox{0.47\textwidth}{!}{%
  \begin{tabular}{lcccc}
    \toprule
    \multicolumn{1}{l}{\textbf{Dataset}} & \multicolumn{1}{c}{\textbf{\# Users}} & \multicolumn{1}{c}{\textbf{\# Items}} & \multicolumn{1}{c}{\textbf{\# Interactions}} & \multicolumn{1}{c}{\textbf{Avg. Length}} \\ 
    \midrule
    ML-1M & $6,040$ & $3,416$ & $999,611$  & $165.5$\\
    Beauty & $22,363$ & $12,101$ & $198,502$  & $8.9$\\
    Video-Games & $24,303$ & $10,673$ & $231,780$ & $9.5$\\
  \bottomrule
\end{tabular}}
\vspace{-0mm}
\end{table}

\noindent \textbf{Baselines.} 
We compare Mamba4Rec against several baseline methods, which include BPR-MF~\cite{rendle2012bpr}, CNN-based Caser~\cite{tang2018personalized}, RNN-based models like GRU4Rec~\cite{hidasi2015session} and NARM~\cite{li2017neural}, Transformer-based models such as SASRec~\cite{kang2018self} and BERT4Rec~\cite{sun2019bert4rec}, and LRURec~\cite{yue2024linear} based on linear recurrent units~\cite{orvieto2023resurrecting}.

\vspace{1.5ex}\noindent \textbf{Evaluation Metrics.}
We adopt Hit Ratio (HR), Normalized Discounted Cumulative Gain (NDCG), and Mean Reciprocal Rank (MRR) with truncation at 10 as the evaluation metrics, \ie, HR@10, NDCG@10, and MRR@10.

\vspace{1.5ex}\noindent \textbf{Implementation Details.}
In the default architecture of our model, a single Mamba layer is employed without position embeddings. The Adam optimizer~\cite{kingma2014adam} is used with a learning rate of 0.001. The training batch size is 2048, and the evaluation batch size is 4096. All models use an embedding dimension of 64. To address the sparsity of the Amazon datasets, a dropout rate of 0.4 is used, compared to 0.2 for MovieLens-1M. The maximum sequence length is set proportionally to the mean number of actions per user: 200 for MovieLens-1M and 50 for Amazon datasets. We follow RecBole~\cite{zhao2021recbole} for further implementation details. For the parameters of the Mamba block, the SSM state expansion factor is 32, the kernel size for 1D convolution is 4, and the block expansion factor for linear projections is 2.

\subsection{Overall Performance}
Table~\ref{tab:performance} demonstrates the overall recommendation performance. Among sequential recommendation baseline models, Transformer-based models (\ie, SASRec, BERT4Rec) and LRU-based models (\ie, LRURec) generally outperform RNN-based models. Notably, by incorporating the selective SSMs, Mamba4Rec generally achieves superior performance compared to all the baselines across sparse and dense datasets, as well as varying maximum sequence lengths, revealing its strength in accurately capturing user preferences and recommending the proper items.
The results highlight the potential of state space models as a compelling alternative to other existing user behavior modeling operators (\eg, RNNs, Transformers, and LRUs) for sequential recommendation.

\subsection{Model Efficiency}

Table~\ref{tab:efficiency} details the efficiency performance of Transformer-based models, LRURec and Mamba4Rec, all measured on a single Nvidia A5000 GPU with 24GB memory. SASRec, BERT4Rec, and LRURec are configured with their default setting of two layers. Considering the tradeoff between efficiency and effectiveness, we use the model with a single Mamba layer instead of models with only a Mamba block or stacked Mamba layers on the long-sequence MovieLens-1M dataset. Mamba4Rec exhibits superior performance in terms of GPU memory usage, training time per epoch, and inference time per batch. Additionally, it demonstrates faster convergence, reaching its best validation performance around epoch 50, compared to LRURec (epoch 55), SASRec (epoch 70), and BERT4Rec (epoch 180). These results suggest that Mamba4Rec achieves superior efficiency and reduces GPU memory cost, particularly on long-sequence datasets when compared to Transformer-based models and LRURec.

\begin{table}[t]
\renewcommand\arraystretch{0.9}
\normalsize
\caption{Efficiency performance on MovieLens-1M}
\vspace{-7pt}
\label{tab:efficiency}
\begin{tabular}{l|c|c|c}
\hline
Method & GPU memory & Training time & Inference time\\
\hline
SASRec  & 14.98GB & 131.24s & 0.27s   \\
BERT4Rec  & 15.48GB & 207.60s & 0.90s   \\
LRURec  & 16.69GB & 319.08s & 0.37s   \\
Mamba4Rec  & 5.01GB & 56.13s & 0.11s     \\
\hline
\end{tabular}
\end{table}

\begin{table}[ht]
	\caption{Ablation analysis on ML-1M and Amazon-Beauty.}
  \vspace{-7pt}
	\label{tab:ablation}
	\begin{center}
	\resizebox{0.47\textwidth}{!}{
	\begin{tabular}{l|cc|cc}
		\toprule
		\multirow{2}{*}{\textbf{Architecture}}&\multicolumn{2}{c}{\textbf{Movielens-1M}}& \multicolumn{2}{c}{\textbf{Amazon-Beauty}} \cr
		\cmidrule(lr){2-3}\cmidrule(lr){4-5}&\textbf{NDCG@10} & \textbf{MRR@10}  & \textbf{NDCG@10} & \textbf{MRR@10} 
            \\ \hline
		Default  & \underline{0.1822} & \underline{0.1425} & \textbf{0.0451} & \textbf{0.0362}	\\
            Block Only & 0.1786 & 0.1404 & 0.0405 & \underline{0.0331} \\
            2 Layers & \textbf{0.1845}  & \textbf{0.1437}  & \underline{0.0440} & 0.0314 \\
            \hline
		w/ PE  & 0.1738 & 0.1342 & 0.0408 & 0.0319  \\	
		w/o PFFN  & 0.1740 & 0.1357 & 0.0437 & 0.0340  \\
            w/o LayerNorm & 0.1820 & 0.1396 & 0.0419 & 0.0335 \\
		\bottomrule
	\end{tabular}}
	\end{center}
  \vspace{-0.2cm}
\end{table}

\subsection{Ablation Study}

To understand the contributions of each component in our architecture, we conduct an ablation study. Table~\ref{tab:ablation} presents the results. We introduce each variant and analyze its respective impact:
\begin{itemize}[leftmargin=10pt]
    \item \textbf{Positional embedding}: The results demonstrate that adding learnable positional embeddings does not improve the performance, due to the recurrent nature of the Mamba block.
    \item \textbf{Layer normalization}: Following each Mamba block and feed-forward network, we employ dropout and layer normalization. Our results demonstrate their effectiveness in mitigating overfitting and enhancing performance.
    \item \textbf{Feed-forward network}: The results indicate that the PFFN layer is particularly beneficial for the longer sequence dataset, MovieLens-1M. This suggests the PFFN's ability to model nonlinearity leads to improvement in performance.
    \item \textbf{Mamba block only}: Replacing the Mamba layer with a Mamba block (and removing PFFN, dropout, and layer normalization) results in faster computation but sacrifices overall performance.
    \item \textbf{Number of Mamba layers}: Stacking two Mamba layers with residual connections yields performance improvements, particularly on the larger, long-sequence MovieLens-1M dataset. This shows the model's ability to handle complex user behavior at the cost of reduced computational efficiency.
\end{itemize}

\section{Conclusion and Future Work}
In this work, we tackle the limitations of Transformer-based sequential recommendation models. We introduce Mamba4Rec built upon the selective structured state space model. Our experiments demonstrate that Mamba4Rec achieves strong performance on datasets with varying sparsity and sequence lengths. Additionally, it has significant improvements in both computational efficiency and memory cost, making it effective and efficient for sequential recommendation tasks.
In the future, we aim to design state space models tailored for recommendation systems and promote the SSM-based models in this domain.

\bibliographystyle{ACM-Reference-Format}
\balance
\bibliography{ref}

\end{document}